# Nonparametric estimation of a mixing distribution for a family of linear stochastic dynamical systems

A. Kryshchenko[a,*], A. Schumitzky[c], M. van Guilder[a], M. Neely[a,b]

[a]*Laboratory of Applied Pharmacokinetics and Bioinformatics, Children's Hospital-Los Angeles, 4650 Sunset Blvd., Los Angeles, CA 90027, USA*

[b]*Pediatric Infectious Diseases, Children's Hospital of Los Angeles, Keck School of Medicine, University of Southern California, Los Angeles, CA 90027, USA*

[c]*Department of Mathematics, University of Southern California, Los Angeles, CA 90089-2532, USA*

**Abstract**

In this paper we develop a nonparametric maximum likelihood estimate of the mixing distribution of the parameters of a linear stochastic dynamical system. This includes, for example, pharmacokinetic population models with process and measurement noise that are linear in the state vector, input vector and the process and measurement noise vectors. Most research in mixing distributions only considers measurement noise. The advantages of the models with process noise are that, in addition to the measurements errors, the uncertainties in the model itself are taken into the account. For example, for deterministic pharmacokinetic models, errors in dose amounts, administration times, and timing of blood samples are typically not included. For linear stochastic models, we use linear Kalman-Bucy filtering to calculate the likelihood of the observations and then employ a nonparametric adaptive grid algorithm to find the nonparametric maximum likelihood estimate of the mixing distribution. We then use the directional derivatives of the estimated mixing distribution to show that the result found attains a global maximum. A simple example using a one compartment pharmacokinetic linear stochastic model is given. In addition to population pharmacokinetics, this research also applies to empirical Bayes estimation.

*Corresponding author. Tel.:+1 323-649-8113; Fax:+1 213-740-2424.

*Email addresses*: akryshchenko@chla.usc.edu (A. Kryshchenko), schumitzky@aol.com (A. Schumitzky), uphill17@att.net (M. van Guilder), mneely@chla.usc.edu (M. Neely),

# 1. Introduction

The mixing distribution problem we consider can be stated as follows. Let $Y_1,...,Y_N$ be a sequence of independent but not necessarily identically distributed random vectors. Each $Y_i$ is a vector of one or more observations from each of $N$ subjects in the population. Let $\theta_1,...,\theta_N$ be sequence of independent and identically distributed random vectors belonging to a subset $\Theta$ of Euclidean space with common but *unknown* distribution $F$. The $\{\theta_i\}$ are not observed. It is assumed that the conditional densities $p(Y_i|\theta_i)$ are known, for $i=1,...,N$. The mixing distribution of $Y_i$ with respect to $F$ is given by

$$p(Y_i|F) = \int p(Y_i|\theta_i) dF(\theta_i).$$

Because of independence of the $\{Y_i\}$, the mixing distribution of the $\{Y_i\}$ with respect to $F$ is given by

(1) $\qquad L(F) = p(Y_1,...,Y_N|F) = \prod_{i=1}^{N} \int p(Y_i|\theta_i) dF(\theta_i)$

*The mixing distribution problem is to maximize $L(F)$ with respect to all distributions $F$ on $\Theta$.*

Note that $L(F)$ is just the likelihood function of the data given $F$. It is important later to note that $L(F)$ is a convex function of $F$. Further, it is shown in (Lindsay, 1983), under simple hypotheses, that the global maximizer $F^{ML}$ of $L(F)$ is a discrete distribution with at most $N$ support points, where $N$ is the number of subjects in the population and a support point is a vector of model parameter values with nonzero probability.

It is common in the literature on mixing distributions to consider a deterministic model of the conditional density $p(Y_i|\theta_i)$, i.e. to consider $Y_i$ to be a function of $\theta_i$ with additive measurement error $v_i$. The measurement noise $v_i$ which is assumed to be normally distributed with mean vector zero and known covariance matrix $V_i(\theta)$. In practice however the model for $p(Y_i|\theta_i)$ is not deterministic as it is affected by the random state-space process of generating $Y_i$. For example, in case of pharmacokinetic problems, errors in the dose amount and timing, so called process noise, are not included in the deterministic models. It is shown in (Jelliffe et al. 1992) that the resulting drug concentrations are heavily influenced by these kinds of errors. The fundamental importance of our paper is that the method we describe is able to account for process and measurement noise in the models. In particular, we consider $Y_i$ to be a vector of discrete measurements for a linear stochastic differential equation, where the state vector includes prosses noise and the measurement vector includes measurement noise.

Once the exact form of the conditional density $p(Y_i|\theta_i)$ has been determined, there are a number of algorithms that can be used for solving the mixing distribution problem, see (Wang 2009) and the references therein. In this paper we use the method of Nonparametric Adaptive Grid (NPAG), see (Leary et al. 2001, Neely et al. 2012, Tatarinova et al. 2013).

**Outline of paper.** This paper is organized as follows: We first discuss the types of models considered based on the form of the conditional densities $\{p(Y_i|\theta_i)\}$. We show that the log likelihood can be reduced to a problem of calculating $p(y_{i(k+1)}|y_{i1},...,y_{ik},\theta)$, for each individual subject $i$. We discuss briefly common simple regression models, which do not allow for the important process noise errors. Then we introduce the main models of interest, where $Y_i$ is the discrete measurement for a stochastic differential equation. These stochastic models accommodate process noise errors. For linear stochastic differential equations, the differential equations can be exactly represented by discrete equations. The likelihood function is defined in terms of a linear Kalman-Bucy filter.

Equally important in this paper is the method for calculating the global optimum $F^{ML}$. This is discussed in Section 4. Our method is different then the popular methods in the literature such as Wang (2007) and Wang (2009). Our method is called Nonparametric Adaptive Grid (NPAG). It is based on modern convex analysis and adaptive discrete optimization. We note that there is a simple condition, which guarantees that a proposed solution $F$ is indeed a global optimum. This is unique to convex optimization.

Finally, we end with an important application of the paper: pharmacokinetic population models. We study a one-compartment model with process and measurement noise and give numerical examples. In particular, we simulate examples with different amount of process and measurement noise and then compare the simulated distributions $F$ to the estimated distributions $F^{ML}$ when we ignore process noise in the model or include it. The results show that simulated distribution $F$ differs from estimated distribution $F^{ML}$ significantly when process noise is not taken into account in the model. This again highlights the main purpose of our paper.

## 2. Models for $p(Y_i | \theta_i)$

The difficulty of the mixing distribution problem is determined by the form of the conditional densities $\{p(Y_i | \theta_i)\}$.

### 2.1 Nonlinear Regression Models

Most of the results in the literature for this mixing distribution problem assume a regression equation of the form

(2)     $Y_i = h_i(\theta_i) + v_i, \ i = 1,...,N$

where $h_i$ is a known vector function and $v_i$ is the normal measurement noise with mean vector zero and known covariance matrix $V_i(\theta)$. In this case $p(Y_i | \theta_i) = \eta(Y_i - H(\theta_i), V_i(\theta))$, where $\eta(Z, \Sigma)$ is the density of the multivatiate normal distribution with mean vector 0, covarince matrix $\Sigma$, evaluated at the vector $Z$.

### 2.2 Stochastic Differential Equation Models

In this paper we consider the mixing distribution problem in a much more complicated setting. It is assumed that the observation vector $Y_i$ is the discrete output of a stochastic differential equation of the form (continuous dynamics, discrete observations):

(3a)     $dx(t) = f(x,u,t,\theta)dt + g(x,u,t,\theta)dw(t), \ t \geq t_0, \ x(t_0) \sim N(\widehat{x_0}(\theta), \Sigma_0(\theta))$

(3b)     $y_k = h_k(x_k, u_k, \theta) + v_k, \ k = 1,...,m$
$Y_i = \{y_1,...,y_m\}$

In Eq. (3a) at time $t$, $x(t)$ is the state vector; $u(t)$ is a known piece-wise continuous input; $\theta \in \Theta$ is a vector of subject-specific parameters for the $i$th subject; $f$ and $h$ are known continuous vector functions; $w(t)$ is a vector white noise process with mean 0 and covariance $E[dw(t), dw(t)^T] = W(t)dt$; and $N(m, S)$ represents the multivatiate normal distribution with mean vector $m$, covarince matrix $S$. In Eq. (3b) at time $t_k$, $y_k$ is the noisy measurement vector; $h_k$ is a known continuous vector function; and $v_k \sim N(0, V_k(\theta))$ is the vector measurement noise.

In the case when $f$, $g$ and $h_k$ are linear functions of their respective arguments, the stochastic system of Eq. (3ab) is called *linear*. Otherwise the system is called *nonlinear*.

## 3. Likelihood Calculations and Kalman-Bucy Filtering

By the telescoping property of conditional densities we have:
$$p(Y_i | \theta) = p(y_{i1}, ..., y_{im} | \theta) = \prod_{k=0}^{m-1} p(y_{i(k+1)} | y_{i1}, ..., y_{im}, \theta)$$
and therefore
$$L(F) = \log p(Y_1, ..., Y_N | F) = \sum_{i=1}^{N} \log(\int \prod_{k=0}^{m-1} p(y_{i(k+1)} | y_{i1}, ..., y_{im}, \theta) dF(\theta))$$

Let $I_{ik} = (y_{i1}, ..., y_{ik}; u_{i1}, ..., u_{ik})$. The crux of the likelihood calculation is in the calculation of $p(y_{i(k+1)} | I_{ik}, \theta)$ for an individual subject.

In the regression case of Eq. (2), $p(y_{i(k+1)} | I_{ik}, \theta) = p(y_{i(k+1)} | \theta)$, and the problem is much simpler. In the general case of Eq. (3), the calculation of $p(y_{i(k+1)} | I_{ik}, \theta)$ is a problem of nonlinear filtering. For the application to population parmacokinetics, Klim et al. (2009) approximate this calculation with the extended Kalman filter. Approximations by particle filtering may be more accurate, see Crisan and Doucer (2002).

### 3.1 Continuous state-discrete observations linear stochastic model

In a later paper we shall address the nonlinear problem. In this paper, we consider only the *linear* stochastic case. Assume we focus on an individual subject. The subscript *i* will be supressed. Now consider Eq. (3), and assume $f$, $g$ and $h$ are linear vector functions. Eq. (3) then becomes

(4a) $\quad dx(t) = A(t,\theta)x(t)dt + B(t,\theta)u(t)dt + dw(t), \quad t \geq 0; \quad x(t_0) \sim N(\widehat{x_0}(\theta), P_0(\theta))$

(4b) $\quad y_k = C_k(\theta)x(t_k) + v_k, \quad k = 1, ..., m$

where $A(t,\theta)$, $B(t,\theta)$ and $C_k(\theta)$ are known continuous matrices. Now assume $u(t)$ is piece-wise constant with $u(t) = u_{k+1}$ on the interval $[t_k, t_{k+1}]$. Then, using the Ito formula, Eq.(4a) can be integrated over the interval $[t_k, t_{k+1}]$ to give an exact discrete time system:

(4c) $\quad x_{k+1} = A_k(\theta)x_k + B_k(\theta)u_k + w_{k+1}(\theta); \quad x(t_0) \sim N(\widehat{x_0}(\theta), \Sigma_0(\theta))$

where $x_k \equiv x(t_k)$; $A_k(\theta) \equiv \Phi(t_{k+1}, t_k, \theta)$ is the fundamental matrix of the homogeneous part of Eq. (3a), $B_k(\theta) = \int_{t_k}^{t_{k+1}} \Phi(t_k, s, \theta) B(s, \theta) ds$; $w_{k+1}(\theta) = \int_{t_k}^{t_{k+1}} \Phi(t_{k+1}, s, \theta) dw(s)$ and $w_{k+1}(\theta)$ is a zero mean Gaussian sequence with the covariance matrix

$$E[w_{k+1}(\theta) w_{k+1}(\theta)'] = \int_{t_k}^{t_{k+1}} \Phi(t_{k+1}, s, \theta) W(s) \Phi'(t_{k+1}, s, \theta) ds,$$ see Jazwinsky (1980, p. 199)

The main theoretical result for calculating $p(y_{k+1} | I^k, \theta)$ in the linear stochastic system of Eq. (4abc) is given by the following Proposition, which is proved in Kumar and Varaiya (1986, Chapter 7, Section 3). Because of its importance, we sketch the proof here.

**Proposition**: *Define*

$$x_{k+1|k}(\theta) \equiv E[x_{k+1} | Y^k] \text{ and } \Sigma_{k+1|k}(\theta) = E[(x_{k+1} - x_{k+1|k})(x_{k+1} - x_{k+1|k})^T | Y^k]$$

*Then the conditional density $p(y_{k+1} | I^k, \theta)$ is normal with mean vector $C_{k+1}(\theta) x_{k+1|k}$ and covariance matrix $C_{k+1}(\theta) \Sigma_{k+1|k} C_{k+1}(\theta)^T + V$.*

Proof. Fix $\theta$. It is also proved in Kumar and Varaiya (1986, Chapter 7, Section 3) that the conditional density $p(x_{k+1} | I^k, \theta)$ is normal, with mean vector $x_{k+1|k}(\theta)$ and covariance matrix $\Sigma_{k+1|k}$. By the property of normal distributions, since $y_{k+1} = C_{k+1}(\theta) x_{k+1} + v_{k+1}$, the conditional density $p(y_{k+1} | I_k, \theta)$ is also normal with mean $E[y_{k+1} | Y^k, \theta] = C_{k+1}(\theta) E[x_{k+1} | Y^k, \theta] = C_{k+1}(\theta) x_{k+1|k}(\theta)$ and covariance

$$E[(y_{k+1} - C_k(\theta) x_{k|k})(y_{k+1} - C_k(\theta) x_{k|k})^T | Y^k, \theta] = C_{k+1}(\theta) \Sigma_{k+1|k}(\theta) C_{k+1}(\theta)^T + V_{k+1} \quad \text{q.e.d.}$$

Note: In Kumar and Varaiya (1986), the above Proposition is proved in the case that the control $u_k$ is a nonlinear feedback function of $I^k$.

**3.2 Kalman-Bucy Linear Filter**

In the above proposition, the terms $x_{k+1|k}$ and $\Sigma_{k+1|k}$ are given recursively by the Kalman-Bucy linear filter:

$x_0 = x(t_0) \sim N(\widehat{x_0}(\theta), P_{0|0}(\theta))$; $x_{k+1|k}(\theta) = A_k(\theta) x_k + B_k(\theta) u_k$;
$\Sigma_{k+1|k}(\theta) = W_k(\theta) + A_k(\theta) \Sigma_k(\theta) (A_k(\theta))^T$; $x_{k+1}(\theta) = x_{k+1|k}(\theta) + F_{k+1}(\theta)[y_{k+1} - C_{k+1}(\theta) x_{k+1|k}(\theta)]$;
$\Sigma_{k+1|k+1}(\theta) = [I - F_{k+1}(\theta)(C_{k+1}(\theta))^T] \Sigma_{k+1|k}(\theta)$, $F_{k+1}(\theta) = \Sigma_{k+1|k}(\theta)(C_{k+1}(\theta))^T (\Sigma_k^y)^{-1}$
$k = 0, ..., K$ see Kumar and Varaiya (1986, pp. 103).
The intitial conditions $\widehat{x_0}(\theta)$ and $\widehat{\Sigma_0}(\theta)$ are supplied by the user.

**4. Optimization of the Likelihood**
Of equal importance in calculating the maximum likelihood estimate is the optimization of the likelihood function in Eq. (1) with respect to $F$.

**4.1 Nonparametric Maximum Likelihood Adaptive Grid Algorithm**
The optimization of Eq.(1) will be done by the nonparametric maximum likelihood adaptive grid (NPAG) algorithm. A brief overview of the NPAG is now given. For complete details see (Yamada et al. 2014). First consider a large grid of fixed supports points $\{\phi_k\}$ on $\Theta$ and let $\lambda = (w_1, ..., w_K)$. The optimization of Eq. (1) is approximated by the maximization of

(5) $\quad l(\lambda) = \sum_{i=1}^{N} \log(\sum_{k=1}^{K} w_k p_i(Y_i | \phi_k))$

with respect to $\{w_k\}$, which is a convex optimization problem. The idea of Robert Leary (at the Pharsight Corporation) and James Burke (at the University of Washington) was to solve this optimization problem by a method consistent with modern convexity theory. Namely, optimize Eq. (5) by the Primal-Dual Interior Point method (IPM) (Boyd and Vandenberghe 2004); see also (Bell 2012).

By convexity theory (see Boyd and Vandenberghe 2004), the IPM algorithm is guaranteed to give a global maximum of $l(\lambda)$ on the specified grid. Notice that the objective function in Eq. (5) depends only on the matrix $\Psi = [p_i(Y_i | \theta_k)]_{\substack{i=1,N \\ k=1, grid\ size}}$

**Adaptive Grid (AG)**

Define an initially large grid G. Calculate $\Psi$ on G. Implement IPM. Remove the support points on G with low probability. Then the NPAG program uses the current solution supports points as a base from which to determine a new expanded adaptive grid (AG). The expanded grid is formed by adding two candidate support points in each dimension of each support point in the old grid. The candidate supports are the vertices of a hypercube centered on each old support point and with segments of length $2\varepsilon(b_k - a_k)$ where $a_k$ and $b_k$ are the minimum and maximum values for each parameter defined by the user and $\varepsilon_j$ is a decreasing sequence of small numbers. Initially $\varepsilon_j$ is set to 0.2. The IPM is applied again and the process repeats with $\varepsilon_{j+1} = \varepsilon_j / 2$. As the algorithm generates better solutions, the size of the hypercube shrinks, resulting in new grid. Since the IPM algorithm is employed at each step of NPAG, the global optimum on current the given grid is guaranteed.

**USC NPAG algorithm**

These are the main steps of the USC NPAG algorithm written in a simple algorithmic form:

Step 1: Initialize grid G
Step 2: Calculate the $\Psi$ matrix on G and call IPM
Step 3: Remove support points on G with low probabilities and renormalize.
 (This completes one cycle)
Step 4: Test exit conditions, i.e. if the difference between successive likelihoods
or the grid diameter is sufficiently small then stop.
If exit conditions are not met, go to Step 2.

**4.2 Directional derivative condition to check for optimality**

One method to check if NPAG has converged to a global maximum is described in (Lindsay 1983). It uses the so-called directional derivative to check if a current distribution $F$ is in fact optimal.
(Convexity theory is unique in this sense that a proposed maximizer can be checked for optimality).

The directional derivative of $F$ in the direction $\theta$ is defined by

(6) $\quad D(\theta, F) = \sum_{i=1}^{N} \frac{p(Y_i | \theta)}{p(Y_i | F)} - N$

**Theorem** (Lindsay 1983): $F^{ML}$ is the distribution that maximizes $L(F)$ in Eq. (1) with respect to all $F$ on $\Theta$ if and only if $\max\{D(\theta, F^{ML}): \theta \in \Theta\} \leq 0$. Moreover the support of $F^{ML}$ is contained in the set of $\theta \in \Theta$ for which the function $D(\theta, F^{ML})$ has a global maximum.

In NPAG, we apply this criterion after the algorithm has converged. In the examples below, we checked that $\max\{D(\theta, F^{ML}): \theta \in \Theta\} \leq 0$, which implies that an optimal solution has been found.

**5. Application to Pharmacokinetic Population Analysis**

Our main application in this paper is to develop a nonparametric maximum likelihood estimate of the distribution of parameters in linear stochastic pharmacokinetic (PK) population models, i.e. PK population models with process noise. The advantages of models with process noise are that, in addition to the measurement errors, additional uncertainties in the data are taken into the account, i.e. process noise. For example, in case of the pharmacokinetic (PK) problem the errors like
- dose errors
- dose timing errors
- sample timing errors

are not included in the deterministic models. The stochastic models on the other hand can accommodate these types of errors.

In a number of papers, e.g. (Klim 2009) and (Mortensen 2007), the authors considered the above maximum likelihood problem for the case where $F$ is assumed to be multivariate normal with unknown mean vector and unknown covariance matrix. For nonlinear models they used the Extended Kalman-Bucy filter to approximately calculate $p(y_{k+1} | I_k, \theta)$. They developed software programs for Matlab and R. More recently, under the same normal hypothesis, Delattre and Lavielle (2011) developed software programs for MONOLIX. We have extended these works to the nonparametric case where $F$ is any probability distribution. This allows $F$ to describe distributions that are multi-modal and long-tailed. This includes the important case of models with genetic phenotypes such as fast and slow metabolizers.

In 1991, Mark Welle wrote an MS Thesis (Welle 1983), under the supervision of Alan Schumitzky, which considered NPML for discrete-time stochastic models. He used the nonparametric EM algorithm (Schumitzky 1991) for optimization of the likelihood. It is known that the EM algorithm is *very* slow. We have extended his work in two ways: 1) we allow linear models defined by stochastic differential equations and 2) we use the general nonparametric adaptive grid (NPAG) algorithm for optimization of the likelihood.

**5.1 One-compartment PK model**

We consider the one-compartment PK model described by the linear stochastic differential equation for the state vector $x(t)$ and discrete time linear equations for the observations $y_k$ at time $t_k$, $k = 1 \ldots m$

(7) $$\frac{dx(t)}{dt} = -Kx(t) + w(t), \quad t \geq t_0,$$

$$x_0 \sim N(\frac{D}{V}, \Sigma_0),$$

$$y_k = x_k + v_k$$

where $K$ is the elimination rate constant; $V$ is the volume of distribution; $D$ is the initial dose of the drug; $w(t)$ is the white Gaussian process with the covariance matrix $W(t)$; $v_k$ is a measurement error that we considered to be normally distributed with zero mean and covariance matrix $V_k$.

We use the Ito integral to integrate the stochastic Eq. (7) over the intervals $[t_k, t_{k+1}]$, see Jazwinski (1970). It follows:

$$x_{k+1} = x_k \exp\{-K(t_{k+1} - t_k)\} + \int_{t_k}^{t_{k+1}} e^{-K(t_{k+1} - t)} dw(t)$$

which can be written as $x_{k+1} = x_k \exp\{-K(t_{k+1} - t_k)\} + w_{k+1}$, where $w_{k+1} = \int_{t_k}^{t_{k+1}} e^{-K(t_{k+1}-t)} dw(t)$ is a white Gaussian sequence with zero mean and the covariance $W_{k+1} = E[w_{k+1} w_{k+1}^T] = e^{-2Kt_{k+1}} \int_{t_k}^{t_{k+1}} e^{2Kt} W(t) dt$.

Note: To evaluate the last integral we need to know $W(t)$. In our example we will assume $W(t)$ is a constant, say $W_c$. Then $W_{k+1} = \frac{W_c}{2K}(1 - e^{-2K(t_{k+1}-t_k)})$ and Eq.(7) can now be written as

$$x_{k+1} = \exp\{-K(t_{k+1} - t_k)\} x_k + w_{k+1},$$

$$x_0 \sim N(\frac{D}{V}, \Sigma_0),$$

(8)  $y_k = x_k + v_k$

## 6. Simulation Results

To illustrate the algorithm just described, we used simulated data. The observations were simulated according to Eq. (8) using the following parameters: $N = 100$, $D = 20$, $t_i = (0.2, 0.4, 0.6, 0.8, 1)$. The volume of distribution is taken to be $Vol_i \sim N(1, 0.2)$. The elimination rate constant $K_i$ is simulated from the mixture of two normal distributions: i.e. $K_i \sim 0.5 N(0.5, 0.05) + 0.5 N(1.5, 0.15)$. The initial state condition is $x_0 \sim N(D/Vol, \Sigma_0)$ where $\Sigma_0 = 0$; and measurement error covariance matrix is $V_k = (0.5)^2$.

### Results

The results below can be classified into the following 4 cases:
a) The data were generated with no process noise and the model did not assume any noise, i.e. $W_c = 0$
b) The data were generated with process noise $W_c = 1$ and the model assumed process noise $W_c = 1$
c) The data were generated with process noise $W_c = 7$ and the model assumed no process noise, i.e. $W_c = 0$
d) The data were generated with process noise $W_c = 7$ and the model assumed process noise $W_c = 7$

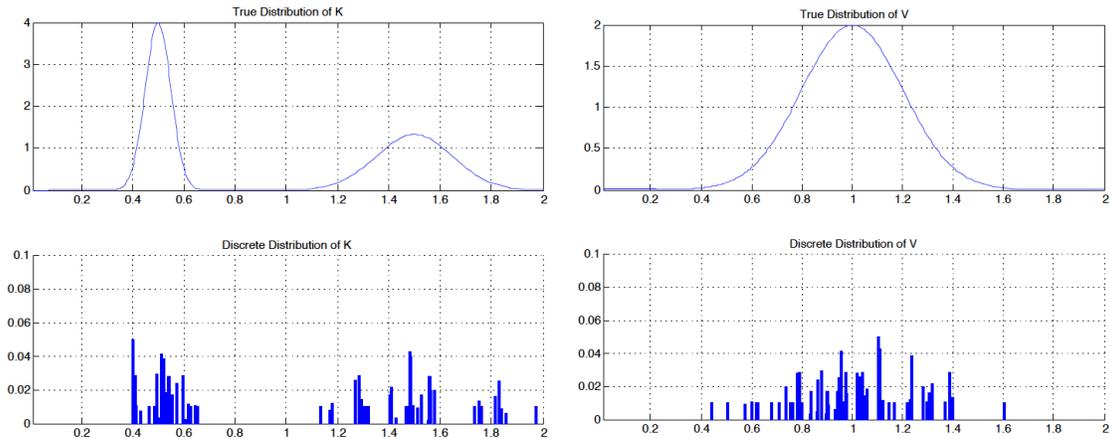

(a)

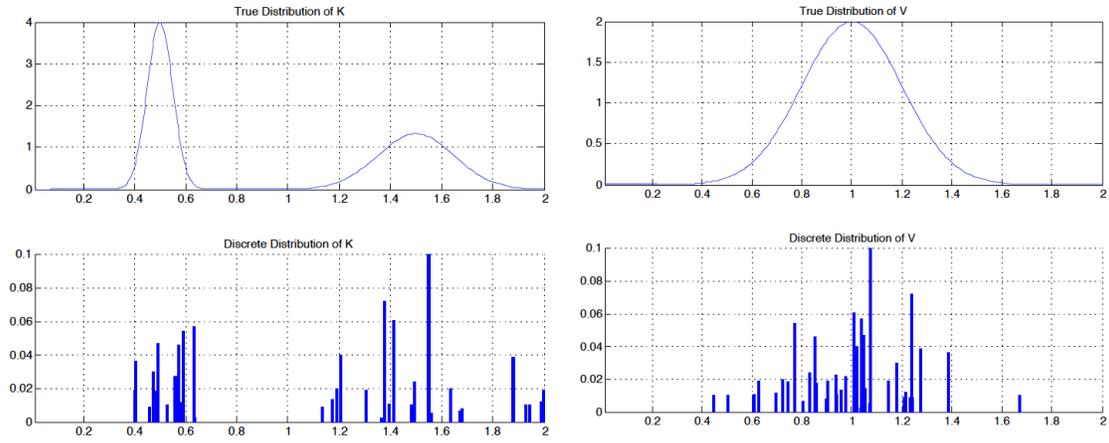

(b)

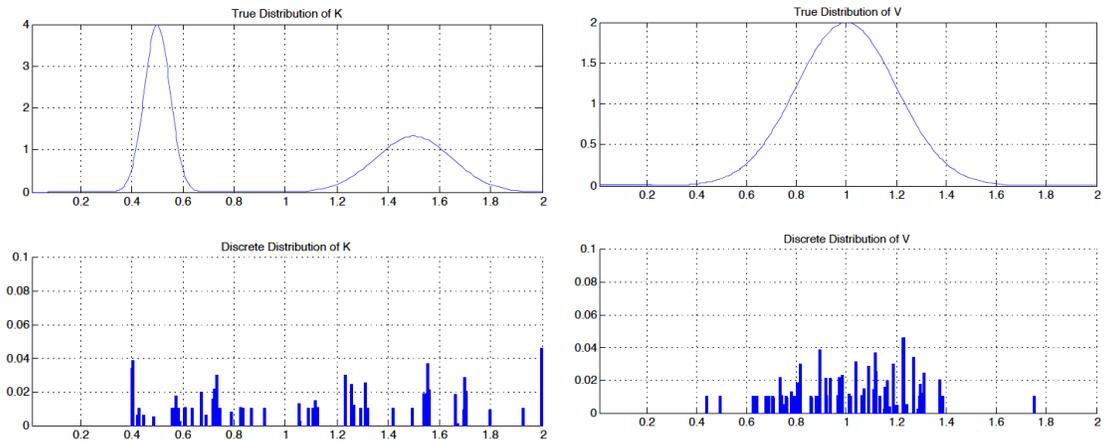

(c)

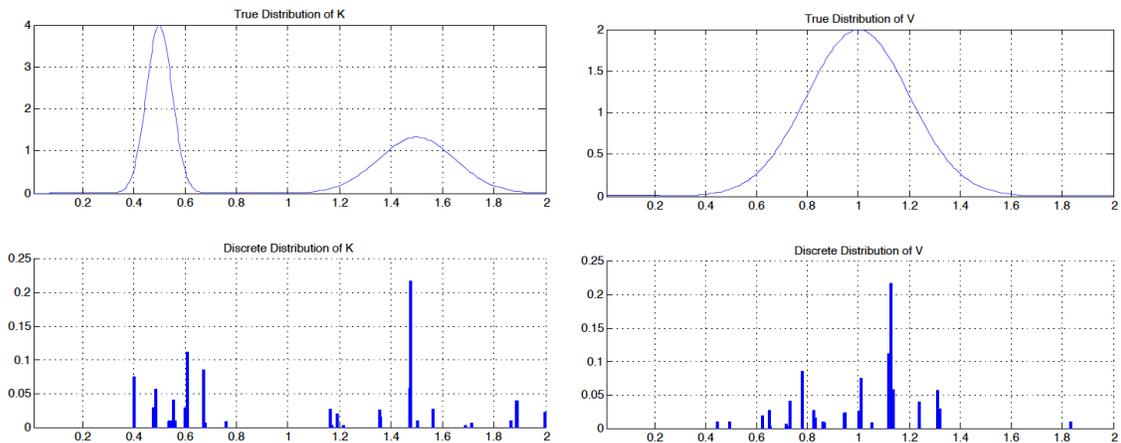

(d)

**Fig. 1** The above figure shows graphs of the true distributions of *K* and *V* (solid lines) and below each graph is the graph of the discrete NPML estimated distribution (spikes).

## 7. Conclusions

In this paper we developed a new algorithm for finding a nonparametric maximum likelihood estimate of the mixing distribution of the parameters of a linear stochastic dynamical system. We demonstrated our algorithm on a one-compartment pharmacokinetic population model with process and measurement noise that is linear in the state vector, input vector and the process and measurement noise vectors. Most research in mixing distributions only considers measurement noise. The advantages of models with process noise are that, in addition to the measurement errors, other uncertainties in the data are taken into account. For example, in the case of deterministic pharmacokinetic models, errors in dose amounts, administration times, and timing of blood samples can be explicitly modeled, separating them from model misspecification. This truly facilitates the appropriate remedial strategy when model predictions are poor, i.e. does the problem lie within the quality of the data or with the structure of the model?

We used linear Kalman-Bucy filtering to calculate the likelihood of the observations and then employed a nonparametric adaptive grid algorithm to find the nonparametric maximum likelihood estimate of the mixing distribution. We then used the directional derivatives of the estimated mixing distribution to show that the result found attained a global maximum. The maximum of the D-function was zero for all four cases, which indicates that the algorithm converged to a global maximum. Note, that in the case of unknown parameter value distributions, e.g. pharmacokinetic data obtained from a clinical study, the modeler must still judge which model is "best" for the data and the circumstances. However, better knowledge of the sources of error, i.e. the data or the model, will facilitate this decision.

When we neglected the correct process noise in the model, we saw that the estimated marginal distributions of $K$ and $V$ were more disperse and chaotic. This was made worse as the simulated process noise was increased. See Fig. 1(c). On the other hand, when the simulated and model process noises were the same, the estimated the marginal distributions of $K$ and $V$ were definitely better, even for large process noise.

We understand, even though being able to account for process noise in mixture models is a significant improvement in the modeling process, most of the real world models are nonlinear. In the future we plan to work on nonlinear stochastic models, which will require nonlinear filtering.

**Acknowledgments**

Support from the National Institutes of Health under grants: GM068968, HD070996 and EB001978 are gratefully acknowledged.

**Appendix**

| Process noise in the data simulation vs. assumed process noise in the program $W_c$ | Resulting weights and support points of $F^{ML}$ | | Log-likelihood Function |
|---|---|---|---|
| | w | (K, V) | |
| | 0.0284 | 0.5936   1.0412 | |
| | 0.0076 | 1.1716   1.2308 | |
| | 0.0155 | 1.2825   0.9776 | |
| | 0.0035 | 0.5369   0.7846 | |
| | 0.0021 | 0.6033   1.0393 | |
| | 0.0260 | 1.2705   1.0283 | |
| | 0.0198 | 1.5778   1.2812 | |
| | 0.0032 | 0.5529   0.8970 | |
| | 0.0167 | 0.5528   0.9007 | |
| | 0.0383 | 0.5225   1.2344 | |
| | 0.0100 | 1.7644   0.6183 | |
| | 0.0501 | 0.4031   1.1025 | |
| | 0.0100 | 0.6277   0.6802 | |
| | 0.0100 | 0.4002   1.1677 | |
| | 0.0288 | 1.2849   0.9743 | |
| | 0.0107 | 0.5448   1.3697 | |
| | 0.0121 | 1.1788   1.2319 | |
| | 0.0100 | 1.1317   0.7828 | |
| | 0.0400 | 1.4852   1.1079 | |
| | 0.0095 | 1.5149   0.9474 | |
| | 0.0166 | 1.5272   0.9404 | |
| | 0.0098 | 1.3171   0.8994 | |
| | 0.0062 | 1.8563   0.9330 | |
| | 0.0029 | 0.5022   0.8773 | |
| | 0.0100 | 1.9760   1.6042 | |
| | 0.0098 | 0.5570   1.1477 | |
| | 0.0082 | 1.4765   1.2217 | |
| | 0.0285 | 0.4111   1.3872 | |
| | 0.0216 | 1.4088   1.3176 | |
| | 0.0015 | 0.4132   1.3907 | |
| | 0.0020 | 1.5525   1.0185 | |
| | 0.0169 | 1.4066   0.8387 | |
| | 0.0100 | 1.7351   0.7992 | |
| | 0.0294 | 0.4965   0.8759 | |
| | 0.0106 | 0.6448   1.0496 | |
| | 0.0106 | 1.4972   0.5996 | |
| | 0.0102 | 1.4764   1.2164 | |
| | 0.0100 | 0.6532   0.9546 | |
| | 0.0185 | 0.5333   1.0575 | |
| | 0.0100 | 1.3136   0.6242 | |
| | 0.0077 | 1.4737   1.1135 | |
| | 0.0282 | 1.5603   1.0174 | |
| | 0.0116 | 0.6179   1.1216 | |
| | 0.0100 | 0.5267   0.4407 | |
| | 0.0284 | 0.5936   1.0412 | |
| | 0.0076 | 1.1716   1.2308 | |
| | 0.0155 | 1.2825   0.9776 | |
| | 0.0035 | 0.5369   0.7846 | |
| | 0.0021 | 0.6033   1.0393 | |
| | 0.0260 | 1.2705   1.0283 | |

| | 1 vs. 1 | w | (K,V) | | -772.5797 |
|---|---|---|---|---|---|
| | | 0.0724 | 1.3772 | 1.2393 | |
| | | 0.0133 | 1.1765 | 0.9587 | |
| | | 0.0460 | 0.5712 | 0.8552 | |
| | | 0.0223 | 0.4896 | 0.9346 | |
| | | 0.0100 | 1.9288 | 0.6043 | |
| | | 0.0463 | 0.4927 | 1.0489 | |
| | | 0.0068 | 1.6703 | 0.8042 | |
| | | 0.0076 | 1.6799 | 0.9004 | |
| | | 0.0190 | 1.3045 | 0.6273 | |
| | | 0.1205 | 1.5502 | 1.0771 | |
| | | 0.0603 | 1.4123 | 1.0087 | |
| | | 0.0088 | 0.4613 | 0.8325 | |
| | | 0.0144 | 0.4859 | 1.0519 | |
| | | 0.0083 | 0.5716 | 1.2351 | |
| | | 0.0191 | 0.4006 | 1.1478 | |
| | | 0.0200 | 1.6378 | 0.7240 | |
| | | 0.0238 | 1.4975 | 0.8309 | |
| | | 0.0114 | 0.5779 | 0.6969 | |
| | | 0.0182 | 0.4878 | 0.7471 | |
| | | 0.0362 | 0.4046 | 1.3879 | |
| | | 0.0055 | 1.5585 | 1.0723 | |
| | | 0.0084 | 0.5639 | 1.2419 | |
| | | 0.0571 | 0.6334 | 1.0379 | |
| | | 0.0110 | 1.3942 | 0.6090 | |
| | | 0.0200 | 1.1937 | 0.7711 | |
| | | 0.0215 | 0.4074 | 0.9742 | |
| | | 0.0191 | 1.9966 | 0.9050 | |
| | | 0.0025 | 1.3645 | 1.2423 | |
| | | 0.0085 | 1.1361 | 1.2094 | |
| | | 0.0021 | 0.6370 | 1.0345 | |
| | | 0.0100 | 0.5281 | 0.4451 | |
| | | 0.0380 | 1.8809 | 1.2732 | |
| | | 0.0121 | 1.9881 | 1.2136 | |
| | | 0.0302 | 0.4758 | 1.1786 | |
| | | 0.0174 | 0.5680 | 0.8601 | |
| | | 0.0541 | 0.5894 | 0.7742 | |
| | | 0.0100 | 1.9445 | 1.6709 | |
| | | 0.0110 | 0.4821 | 0.9409 | |
| | | 0.0272 | 0.5580 | 1.2380 | |
| | | 0.0100 | 1.4799 | 0.5041 | |
| | | 0.0395 | 1.2069 | 1.0195 | |
| | | w | (K,V) | | |
| | | 0.0215 | 0.7210 | 0.7347 | |
| | | 0.0176 | 0.5729 | 1.2977 | |
| | | 0.0114 | 0.4021 | 1.0396 | |
| | | 0.0139 | 0.4000 | 1.1112 | |
| | | 0.0186 | 1.6611 | 0.8019 | |
| | | 0.0100 | 1.1063 | 0.6353 | |
| | | 0.0120 | 1.2673 | 1.3009 | |
| | | 0.0381 | 0.4044 | 0.8931 | |
| | | 0.0102 | 0.5737 | 0.6898 | |
| | | 0.0127 | 1.0525 | 1.1548 | |
| | | 0.0100 | 1.3200 | 0.6391 | |
| | | 0.0103 | 0.8362 | 1.0170 | |
| | | 0.0215 | 0.7210 | 0.7347 | |
| | | 0.0176 | 0.5729 | 1.2977 | |
| | | 0.0114 | 0.4021 | 1.0396 | |
| | | 0.0139 | 0.4000 | 1.1112 | |
| | | 0.0186 | 1.6611 | 0.8019 | |
| | | 0.0100 | 1.1063 | 0.6353 | |

| | 7 vs. 7 | w | (K,V) | | -977.2933 |
|---|---|---|---|---|---|
| | | 0.0228 | 0.4014 | 0.9499 | |
| | | 0.0267 | 1.1633 | 0.6515 | |
| | | 0.1108 | 0.6086 | 1.1200 | |
| | | 0.0097 | 0.5419 | 0.8642 | |
| | | 0.0101 | 1.8672 | 1.8347 | |
| | | 0.0148 | 1.3618 | 1.0005 | |
| | | 0.0182 | 1.9993 | 0.6233 | |
| | | 0.0060 | 0.6780 | 0.7797 | |
| | | 0.0100 | 1.5045 | 0.4966 | |
| | | 0.0062 | 1.7121 | 0.7170 | |
| | | 0.0250 | 1.3568 | 1.0028 | |
| | | 0.0196 | 1.1938 | 0.8270 | |
| | | 0.0100 | 0.5587 | 0.4442 | |
| | | 0.0263 | 1.5632 | 0.8276 | |
| | | 0.0146 | 1.5614 | 0.8305 | |
| | | 0.0853 | 0.6742 | 0.7809 | |
| | | 0.0021 | 1.6884 | 0.7259 | |
| | | 0.0554 | 0.4863 | 1.3111 | |
| | | 0.0281 | 0.4762 | 1.3182 | |
| | | 0.0585 | 1.4728 | 1.1365 | |
| | | 0.2171 | 1.4785 | 1.1287 | |
| | | 0.0071 | 0.5388 | 0.8651 | |
| | | 0.0219 | 1.9998 | 0.9430 | |
| | | 0.0027 | 1.1722 | 0.6524 | |
| | | 0.0383 | 1.8873 | 1.2380 | |
| | | 0.0276 | 0.5991 | 1.1213 | |
| | | 0.0750 | 0.4028 | 1.0139 | |
| | | 0.0069 | 0.7570 | 1.0527 | |
| | | 0.0032 | 1.2162 | 0.8249 | |
| | | 0.0398 | 0.5540 | 0.7325 | |

**Table 1**: This table shows the results of NPAG that were calculated using different levels of process noise. It also compares the results when the process noise was included in the simulation but was not present in the model.